\documentclass[
 reprint,
 amsmath,amssymb,
 aps,
]{revtex4-1}

\usepackage{graphicx}
\usepackage{dcolumn}
\usepackage{mathrsfs}
\usepackage{amsmath}
\usepackage{bm}
\usepackage{color}
\usepackage{float}
\usepackage{CJK}

\begin{document}

\preprint{APS/123-QED}

\title{The restriction of $\sigma^{*}$ and $\phi$ on the protoneutron stars PSR J0348+0432}
\author{Bin Hong}
\affiliation{
Department of Physics, Nanjing University, Nanjing 210008, People's Republic of China
}%
\author{Zhong-Zhou Ren}
 \email{zren@nju.edu.cn}
\affiliation{
Department of Physics, Nanjing University, Nanjing 210008, People's Republic of China and
\\
Department of Physics, Tongji University, Shanghai 200092, People's Republic of China
}%




\date{\today}

\begin{abstract}
The restriction of $\sigma^{*}$ and $\phi$ on the protoneutron star (PNS) PSR J0348+0432 is described by the relativistic mean field theory (RMFT) through choosing the effective coupling constants. We use an entropy per baryon S=1 to describe thermal effect on PSR J0348+0432 in this work and compare the differences between PNS PSR J0348+0432 with $\sigma^{*}$,$\phi$ and without $\sigma^{*}$,$\phi$. These effects include the particle number distribution, the mass-radius relation, the moment of inertia and the surface gravitational redshift. The PNS PSR J0348+0432 with $\sigma^{*}$ and $\phi$ has more nucleons and will pushed forward the threshold for the appearance of the hyperons. The mass-radius relations are ($2.010M_{\odot}$,12.6520km) with $\sigma^{*}$ and $\phi$ and ($2.010M_{\odot}$,12.6170km) without $\sigma^{*}$ and $\phi$ respectively. The moment of inertia corresponding PNS PSR J0348+0432 are ($2.010M_{\odot}$,1.510$\times 10^{45}$g.cm$^{2}$) and ($2.010M_{\odot}$,1.559$\times 10^{45}$g.cm$^{2}$) respectively, the surface gravitational redshift corresponding PNS PSR J0348+0432 are ($2.010M_{\odot}$,0.3747) and ($2.010M_{\odot}$,0.3701) respectively. With the help of these calculations, we study the restriction of $\sigma^{*}$ and $\phi$ on the interactions between baryons in PNS core.

\begin{description}
\item[PACS numbers]
26.60.-c, 26.60.Kp, 21.65.Jk, 24.10.Pa 
\end{description}
\end{abstract}

\maketitle


\section{Introduction}
Neutron stars, are the high-density stars and give a extreme physical condition. Recently, the gravitational-wave event GW170817 was detected from a binary neutron star inspiral puts the neutron stars into a hot research topic ~\cite{01,02}. It is quite meaningful to study theoretically with various methods, especially for the massive neutron stars PSR J1614-2230 whose mass is 1.97$\pm$0.04$M_{\odot}$ using the method of shapiro delay by Demorest et al in 2010 ~\cite{03} and PSR J0348+0432 whose mass is 2.01$\pm$0.04$M_{\odot}$  measured by a combination of radio timing and precise spectroscopy of white dwarf companion by Antoniadis et al in 2013 ~\cite{04}. So far, only these two massive neutron stars whose mass exceed 2$M_{\odot}$ have been observed and thus poses the tightest reliable lower bound on the maximum mass of neutron star. Some theoretical calculations and analytical approaches have been identified to support these observations ~\cite{05,06,07,08}.

Describing a neutron star under the relativistic mean field (RMFT) theory have been adopted in many studies ~\cite{09,10,11}, which consider the baryon-baryon interaction through the exchange of $\sigma,\rho,\omega$ mesons. However, if we consider an additional interaction which couples strongly to strange baryons, the $\sigma^{*}$ and $\phi$ mesons should be included ~\cite{12}.

Lots of work discussed the neutron stars based on zero temperature ~\cite{13,14,15}. For example, the work by Xian-Feng Zhao ~\cite{16} discuss the effect of $\sigma^{*}$ and $\phi$ on the surface gravitational redshift of PSR J0348+0432 and give a meaningful result, however, the result was obtained under the zero temperature. As we know, a cold neutron star is a kind of evolutionary outcome of a PNS which is formed after enormous supernova. So the properties and structure of a PNS should be promoted, but little attentions have been on this topic.

In this paper, with strangeness-rich hyperons in neutron star interior. Under the relativistic mean field (RMF) theory, we investigate the influence of $\sigma^{*}$ and $\phi$ mesons on massive neutron star PSR J0348+0432 at finite entropy. The paper is organized by follows. In Sec.2, we give the complete form of relativistic mean field of hadron interaction at finite entropy including the $\sigma^{*}$ and $\phi$ mesons. In Sec.3, the details in discussion about selecting hyperon coupling constants. In Sec.4, some calculation results of the $\sigma^{*}$ and $\phi$ mesons effect on massive PNS PSR J0348+0432. In Sec.5, a summary is presented.

\section{The RMFT AT FINITE ENTROPY}
 The relativistic mean field (RMF) theory is an effective field theory dealing with hadron-hadron interactions ~\cite{15,16,17,18,19,20}. The degrees of freedom relevant to this theory are baryons interacting through the exchange of $\sigma,\omega,\rho$ mesons, of which the scalar meson $\sigma$ provides the medium-range attraction, the vector meson $\omega$ provides short-range repulsion, and the vector - isospin vector meson $\rho$ describes the difference between neutrons and protons.

The partition function of system is the starting point to study the thermal neutron stars in RMF. From the partition function we could get various thermodynamic quantities at equilibrium.

 For the grand canonical ensemble, the partition function can be written as:
\begin{equation}
  Z=Tr \{ exp[-(\hat{H}-\mu \hat{N})/T] \},
\end{equation}
where $\hat{H}$ denotes the Hamiltonian operator, $\hat{N}$ denotes represents the particle operator, $\mu$ and $T$ denotes the chemical potential and the temperature respectively. We can get the particle population density, the energy density and pressure from the partition function:
\begin{eqnarray}
   n&=&\frac{T}{V} \frac{\partial lnZ}{\partial \mu},\\
  \varepsilon&=&\frac{T^{2}}{V}\frac{\partial lnZ}{\partial T}+\mu n,
  \\
  P&=&\frac{T}{V} lnZ,
\end{eqnarray}
here, $V$ is the volume.
Considering the baryons $B$ and leptons $l$ as fermions, we can get:
 \begin{eqnarray}\nonumber
 lnZ_{B,l}&=&\sum_{B,l} \frac{2J_{B,l}+1}{2\pi^2}\int_{0}^{\infty }k^2dk \{ln[1+e^{-(\varepsilon_{B,l}(k)-\mu_{B,l})/T}]\\
 &+&\frac{V}{T}<\mathcal{L}>,
\end{eqnarray}
The spin quantum number is represented by $J_{B,l}$ and the chemical potential of baryon and lepton is represented by $\mu_{B,l}$. $\mathcal{L}$ denotes the Lagrangian density. $\varepsilon_{B,l}(k)=\sqrt{k^2+m_{B,l}^2}$ denotes thermal excitation energy of baryon and lepton.

The total partition function $Z_{total}=Z_{B}Z_{l}$, $Z_{B}$ and $Z_{l}$ denote the partition function of baryons and the standard noninteracting partition function of leptons respectively. The additional condition of charge neutrality equilibrium is listed as following:
\begin{eqnarray}
\nonumber
  &\sum_{B,l}&\frac{2J_{B,l}+1}{2\pi^{2}}q_{B,l}\int^{\infty}_{0}k^{2}n_{B,l}(k)dk=0,
\end{eqnarray}
where $n_{B}(k)$ and $n_{l}(k)$ denote Fermi distribution function of baryons and leptons respectively. They are represented by
\begin{eqnarray}
  n_{i}(k)=\frac{1}{1+exp[(\varepsilon_{i}(k)-\mu_{i})/T]}(i=B,l).
\end{eqnarray}
When neutrinos are not trapped, the set of equilibrium chemical potential relations required by the general condition:
\begin{eqnarray}
 \mu_{i}=b_{i}\mu_{n}-q_{i}\mu_{e},
\end{eqnarray}
where $b_{i}$ denotes the baryon number of particle $i$ and $q_{i}$ denotes its charge.

The properties of neutron star at finite temperature can be described by the entropy per baryon, the total entropy per baryon is calculated using $S=(S_{B}+S_{l})/(T\rho_{B})$, where $S_{B}=P_{B}+\varepsilon_{B}-\sum_{i=B}\mu_{i}\rho_{i}$ and $S_{l}=P_{l}+\varepsilon_{l}-\sum_{i=l}\mu_{i}\rho_{i}$ ~\cite{21}.

The Lagrangian density of hadron matter is given by ~\cite{22}:
\begin{eqnarray}\nonumber
\mathcal{L}&=&
\sum_{B}\overline{\Psi}_{B}(i\gamma_{\mu}\partial^{\mu}-{m}_{B}+g_{\sigma B}\sigma-g_{\omega B}\gamma_{\mu}\omega^{\mu}
\nonumber\\
&&-\frac{1}{2}g_{\rho B}\gamma_{\mu}\tau\cdot\rho^{\mu})\Psi_{B}+\frac{1}{2}\left(\partial_{\mu}\sigma\partial^{\mu}\sigma-m_{\sigma}^{2}\sigma^{2}\right)
\nonumber\\
&&-\frac{1}{4}\omega_{\mu \nu}\omega^{\mu \nu}+\frac{1}{2}m_{\omega}^{2}\omega_{\mu}\omega^{\mu}-\frac{1}{4}\rho_{\mu \nu}\cdot\rho^{\mu \nu}+\frac{1}{2}m_{\rho}^{2}\rho_{\mu}\cdot\rho^\mu
\nonumber\\
&&-\frac{1}{3}g_{2}\sigma^{3}-\frac{1}{4}g_{3}\sigma^{4}+\sum_{l=e,\mu}\overline{\Psi}_{l}\left(i\gamma_{\mu}\partial^{\mu}
-m_{l}\right)\Psi_{l}
.\
\end{eqnarray}
where the sum on $B$ runs over the octet baryons $(n,p,\Lambda,\Sigma^{-},\Sigma^{0},\Sigma^{+},\Xi^{-},\Xi^{0})$, and $\Psi_{B}$ is the baryon field operator. The last term represents the free lepton Lagrangian. In this work, an additional scalar meson $\sigma^{*}$ and a vector meson $\phi$ are considered by us, their interaction among hyperons need to know, which can be described by the Lagrangian density $\mathcal{L'}$ ~\cite{23}
\begin{eqnarray}\nonumber
\mathcal{L'}&=&
\sum_{B}\overline{\Psi}_{B}(g_{\sigma^{*}B}\sigma^{*}-g_{\phi B}\gamma_{\mu}\phi^{\mu})\Psi_{B}
\nonumber\\
&&+\frac{1}{2}(\partial_{\mu}\sigma^{*}\partial^{\mu}\sigma^{*}-m_{\sigma^{*}}^{2}\sigma^{*2})
\nonumber\\
&&-\frac{1}{4}\phi_{\mu\nu}\phi^{\mu\nu}+\frac{1}{2}m_{\phi}^{2}\phi_{\mu}\phi^{\mu}
\end{eqnarray}
The formula of energy density and pressure of a neutron star at finite temperature under the relativistic mean field theory are given as follows:
\begin{widetext}
\begin{eqnarray}\nonumber
\mathbf{\varepsilon}&=&\frac{1}{3}g_{2}\sigma^{3}+\frac{1}{4}g_{3}\sigma^{4}+\frac{1}{2}m_{\sigma}^{2}\sigma^{2}+\frac{1}{2}m_{\omega}^{2}\omega_{0}^{2}+\frac{1}{2}m_{\rho}^{2}\rho_{03}^{2}\\
\nonumber
&+&\sum_{B}\frac{2J_{B}+1}{2\pi^{2}}\int_{0}^{\infty}\sqrt{k^{2}+(m^{\ast})^{2}}(exp[(\varepsilon_{B}(k)-\mu_{B})/T]+1)^{-1}k^{2}dk\\
&+&\sum_{l}\frac{2J_{l}+1}{2\pi^{2}}\int_{0}^{\infty}\sqrt{k^{2}+m_{l}^{2}}(exp[(\varepsilon_{l}(k)-\mu_{l})/T]+1)^{-1}k^{2}dk,\\
\nonumber
P&=&-\frac{1}{3}g_{2}\sigma^{3}-\frac{1}{4}g_{3}\sigma^{4}-\frac{1}{2}m_{\sigma}^{2}\sigma^{2}+\frac{1}{2}m_{\omega}^{2}\omega_{0}^{2}+\frac{1}{2}m_{\rho}^{2}\rho_{03}^{2}\\
\nonumber
&+&\frac{1}{3}\sum_{B}\frac{2J_{B}+1}{2\pi^{2}}\int_{0}^{\infty}\frac{k^{2}}{\sqrt{k^{2}+(m^{\ast})^{2}}}\left(exp[(\varepsilon_{B}(k)-\mu_{B})/T]+1\right)^{-1}k^{2}dk\\
&+&\frac{1}{3}\sum_{l}\frac{2J_{l}+1}{2\pi^{2}}\int_{0}^{\infty}\frac{k^{2}}{\sqrt{k^{2}+m_{l}^{2}}}(exp[(\varepsilon_{l}(k)-\mu_{l})/T]+1)^{-1}k^{2}dk,
\end{eqnarray}
\end{widetext}
where, $m^{*}=m_{B}-g_{\sigma B}\sigma$ denotes the effective mass of baryon. $B$ and $l$ denote baryons and leptons respectively.

Once the equation of state is specified, the mass and radius of neutron star can be obtained by solving the well-known hydrostatic equilibrium equations of Tolman- Oppenheimer-Volkoff~\cite{24}.
\begin{eqnarray}
\frac{\mathrm dp}{\mathrm dr}&=&-\frac{\left(p+\varepsilon\right)\left(M+4\pi r^{3}p\right)}{r \left(r-2M \right)},
\\\
M(r)&=&4\pi\int_{0}^{r}\varepsilon r^{2}\mathrm dr
.\
\end{eqnarray}

In a uniformly slow-rotating and axially symmetric neutron star, the moment of inertia is given by the following expression \cite{25}:
\begin{eqnarray}
\nonumber
  I\equiv\frac{J}{\Omega}
  &=&\frac{8\pi}{3}\int^{R}_{0}r^{4}e^{-\nu(r)}\frac{\bar{\omega}(r)}{\Omega}\\
  &\times&\frac{(\varepsilon(r)+P(r))}{\sqrt{1-2GM(r)/r}}dr,
\end{eqnarray}
where $J$ denotes the angular momentum, $\Omega$ denotes the angular velocity of the star, $\nu(r)$ and $\bar{\omega}(r)$ denotes radially dependent metric functions, and $R, M(r), \varepsilon(r)$ and $P(r)$ denotes the radius, mass, energy density and pressure of the star respectively. The specific form of $\nu(r)$ is determined by the following expression:
\begin{eqnarray}
\nonumber
  \nu(r)&=&-G\int^{R}_{r}\frac{(M(r)+4\pi x^{3}P(x))}{x^{2}(1-2GM(x)/x)}dx\\
  &+&\frac{1}{2}ln\left(1-\frac{2GM}{R}\right).
 \end{eqnarray}
In particular, the dimensionless relative frequency $\tilde{\omega}(r)\equiv\bar{\omega}(r)/\Omega$ satisfies the following second-order differential equation:
\begin{eqnarray}
  \frac{d}{dr}\left(r^{4}j(r)\frac{d\widetilde{\omega}(r)}{dr}\right)+4r^{3}\frac{dj(r)}{dr}\widetilde{\omega}(r)=0,
\end{eqnarray}
where
\begin{eqnarray}
j(r)=
\begin{cases}
  e^{-\nu(r)}\sqrt{1-2GM(r)/r} & r \leqslant R, \\
  1 & r > R.
\end{cases}
\end{eqnarray}
Note that $\widetilde{\omega}(r)$ is subject to the following two boundary conditions:
\begin{eqnarray}
\nonumber
\widetilde{\omega}'(0)=0,\\
\widetilde{\omega}(R)+\frac{R}{3}\widetilde{\omega}'(R)=1.
\end{eqnarray}
Combining with the EOS and the OV equation, Eqs. (14-18) will be solved.

General relativity gives the gravitational redshift of the star satisfied the relation \cite{26,27}:
\begin{eqnarray}
  z=\left(1-\frac{2GM}{c^{2}R}\right)^{-1/2}-1,
\end{eqnarray}
where $M, R$ denote the mass and radius of the neutron star respectively.
\section{Coupling parameters}
Among the coupling constants for the RMF models, the nucleon coupling constants in the vicinity of the saturation properties of nuclear matter can be determined, such as nuclear saturation density, binding energy per baryon number, effective mass of the nucleon, nuclear compression modulus and asymmetry energy coefficient~\cite{28}. In this study, we choose the parameter set GL85 and the parameter set GL97 listed in Tables~\ref{tab:table1} and Tables~\ref{tab:table2}. These two parameters are often adopted and may well describe the interaction between nucleons ~\cite{22}.
\begin{table}[!htb]
\caption{\label{tab:table1}
 GL85 nucleon coupling constants.
}
\begin{ruledtabular}
\begin{tabular}{ccccccccc}
&$m$&$m_{\sigma}$ &$m_{\omega}$&$m_{\rho}$&$g_{\sigma}$&$g_{\omega}$&$g_{\rho}$&$g_{2}$\\
&$\mathrm{MeV}$&$\mathrm{MeV}$&$\mathrm{MeV}$&$\mathrm{MeV}$& & & &$\mathrm{fm^{-1}}$\\
\hline
&939 &500          &782         &770       & 7.9955    &9.1698 &9.7163&10.07\\
&$g_{3}$&$\rho_{0}$&$B/A$&$K$&$a_{sym}$&$m^{*}/m$&\\
 & &$\mathrm{fm^{-3}}$ &$\mathrm{MeV}$ &$\mathrm{MeV}$ &$\mathrm{MeV}$ &\\
\hline
&29.262    &0.145         &15.95       & 285    &36.8 &0.77&\\
\end{tabular}
\end{ruledtabular}
\end{table}

\begin{table}[!htb]
\caption{\label{tab:table2}
 GL97 nucleon coupling constants.
}
\begin{ruledtabular}
\begin{tabular}{ccccccccc}
&$m$&$m_{\sigma}$ &$m_{\omega}$&$m_{\rho}$&$g_{\sigma}$&$g_{\omega}$&$g_{\rho}$&$g_{2}$\\
&$\mathrm{MeV}$&$\mathrm{MeV}$&$\mathrm{MeV}$&$\mathrm{MeV}$& & & &$\mathrm{fm^{-1}}$\\
\hline
&939 &500          &782         &770       & 7.9835    &8.7 &8.5411 &20.966\\
&$g_{3}$&$\rho_{0}$&$B/A$&$K$&$a_{sym}$&$m^{*}/m$&\\
 & &$\mathrm{fm^{-3}}$ &$\mathrm{MeV}$ &$\mathrm{MeV}$ &$\mathrm{MeV}$ &\\
\hline
&-9.835    &0.153         &16.3       & 240    &32.5 &0.78&\\
\end{tabular}
\end{ruledtabular}
\end{table}
When hyperons are included, their coupling constants are needed. For the coupling constants related with hyperons, we define the ratios:
\begin{eqnarray}
x_{\sigma H}&=&\frac{g_{\sigma H}}{g_{\sigma}}=x_{\sigma}
\\
x_{\omega H}&=&\frac{g_{\omega H}}{g_{\omega}}=x_{\omega}
\\
x_{\rho H}&=&\frac{g_{\rho H}}{g_{\rho}}=x_{\rho}
\
\end{eqnarray}
Where $H$ denotes hyperons ($\Lambda, \Sigma$ and $\Xi$). The ratios of hyperon coupling constant to nucleon coupling constant exist considerable uncertainty. It cannot be decided by the saturation properties of nuclear matter, but could be extrapolated through the hypernuclear experimental data. The hypernuclear potential depth in nuclear matter $U^{N}_{H}$, which is known in accordance with available hypernuclear data, serves to strictly correlate the value of $x_{\sigma H}$ and $x_{\omega H}$~\cite{29}:
\begin{eqnarray}
  U^{N}_{H}=x_{\omega H}V-x_{\sigma H}S,
\end{eqnarray}
where $S=m-m^{*}$, $V=(g_{\omega}/m_{\omega})^{2}\rho_{0}$ are the values of scalar and vector field strengths for symmetric nuclear matter at saturation respectively. With $U^{N}_{H}$, if we give the value of $x_{\omega H}$,we can get the value of $x_{\sigma H}$. The experimental data of hypernuclear potential depth of $U^{N}_{\Lambda}, U^{N}_{\Sigma}$ and $U^{N}_{\Xi}$ are~\cite{30,31,32,33,34,35}:
\begin{eqnarray}\nonumber
U^{N}_{\Lambda}&=&-30\mathrm{MeV},\\
\nonumber
U^{N}_{\Sigma}&=&+30\mathrm{MeV},\\
U^{N}_{\Xi}&=&-15\mathrm{MeV}.
\end{eqnarray}

\begin{figure}[!htb]
\begin{center}
\includegraphics[width=3.5in]{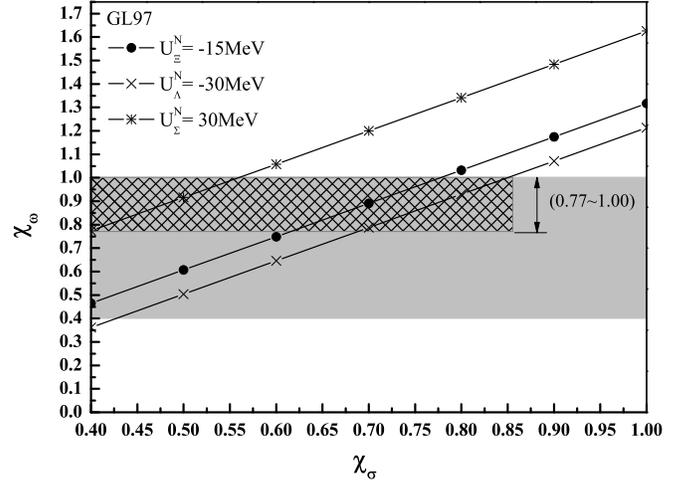}
\caption{The relation of $x_{\sigma}$ and $x_{\omega}$ according to the the hypernuclear potential depth in nuclear matter $U^{N}_{H}$ with GL97.}
\label{figure1}
\end{center}
\end{figure}

\begin{figure}[!htb]
\begin{center}
\includegraphics[width=3.5in]{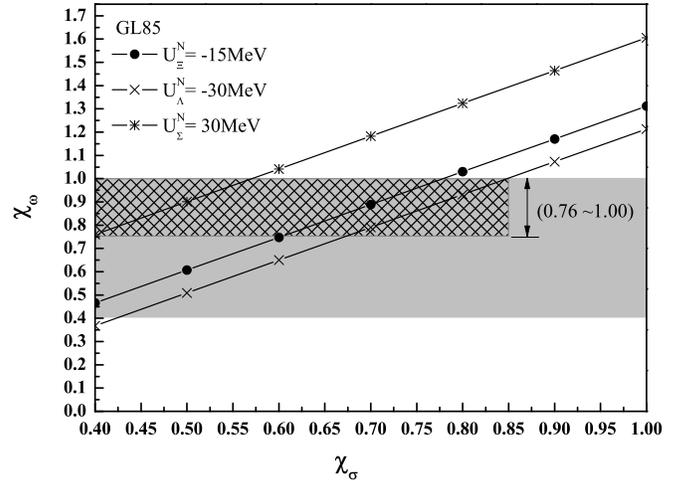}
\caption{The relation of $x_{\sigma}$ and $x_{\omega}$ according to the the hypernuclear potential depth in nuclear matter $U^{N}_{H}$ with GL85.}
\label{figure2}
\end{center}
\end{figure}

In studying the properties of a neutron star with RMF theory, due to the considerable uncertainty in the value of $x_{\omega H}$. The reference~\cite{36} points that its value should be restricted at $1/3$ to $1$.
In this paper, we select $x_{\sigma}$=0.4, 0.5, 0.6, 0.7, 0.8, 0.9, 1.0. For each $x_{\sigma}$, the $x_{\omega}$ will be obtained according to the the hypernuclear potential depth in nuclear matter $U^{N}_{H}$, which are listed in Fig.~\ref{figure1} and Fig.~\ref{figure2}. The calculations point that the the $x_{\omega}$ should be restricted at a narrow area depicted by grid.

 The hyperon coupling constants $x_{\rho\Lambda}, x_{\rho\Sigma}$ and $x_{\sigma\Xi}$ are determined by using SU(6) symmetry\cite{37}:
 \begin{eqnarray}
   x_{\rho\Lambda}=0, x_{\rho\Sigma}=2, x_{\sigma\Xi}=1.
 \end{eqnarray}

The parameters between hyperon-hyperon through exchanging the strange scalar meson ($\sigma^{*}$) and strange vector meson ($\phi$) could be selected as follows.

For the vector meson $\phi$, according to the quark model relationships, the coupling parameters yield to $2g_{\phi\Lambda}=2g_{\phi\Sigma}=g_{\phi\Xi}=-\frac{2\sqrt{2}}{3}g_{\omega} $.

For the scalar meson $\sigma^{*}$, we use the mass of the observed $\sigma^{*}(975)$ meson, but treat its coupling purely phenomenologically so as to satisfy the potential depths $U_{\Sigma}^{(\Xi)}\approx U_{\Lambda}^{(\Xi)}\approx U_{\Xi}^{(\Xi)} \approx 2U_{\Lambda}^{(\Lambda)}\approx 2U_{\Sigma}^{(\Lambda)}=40MeV$. This yield $g_{\sigma^{*}\Lambda}/g_{\sigma}=g_{\sigma^{*}\Sigma}/g_{\sigma}=0.69,g_{\sigma^{*}\Xi}/g_{\sigma}=1.25$~\cite{12}.

\section{DISCUSSION}
\subsection{MASS AND RADIUS}
\begin{figure}[!htb]
\begin{center}
\includegraphics[width=3.5in]{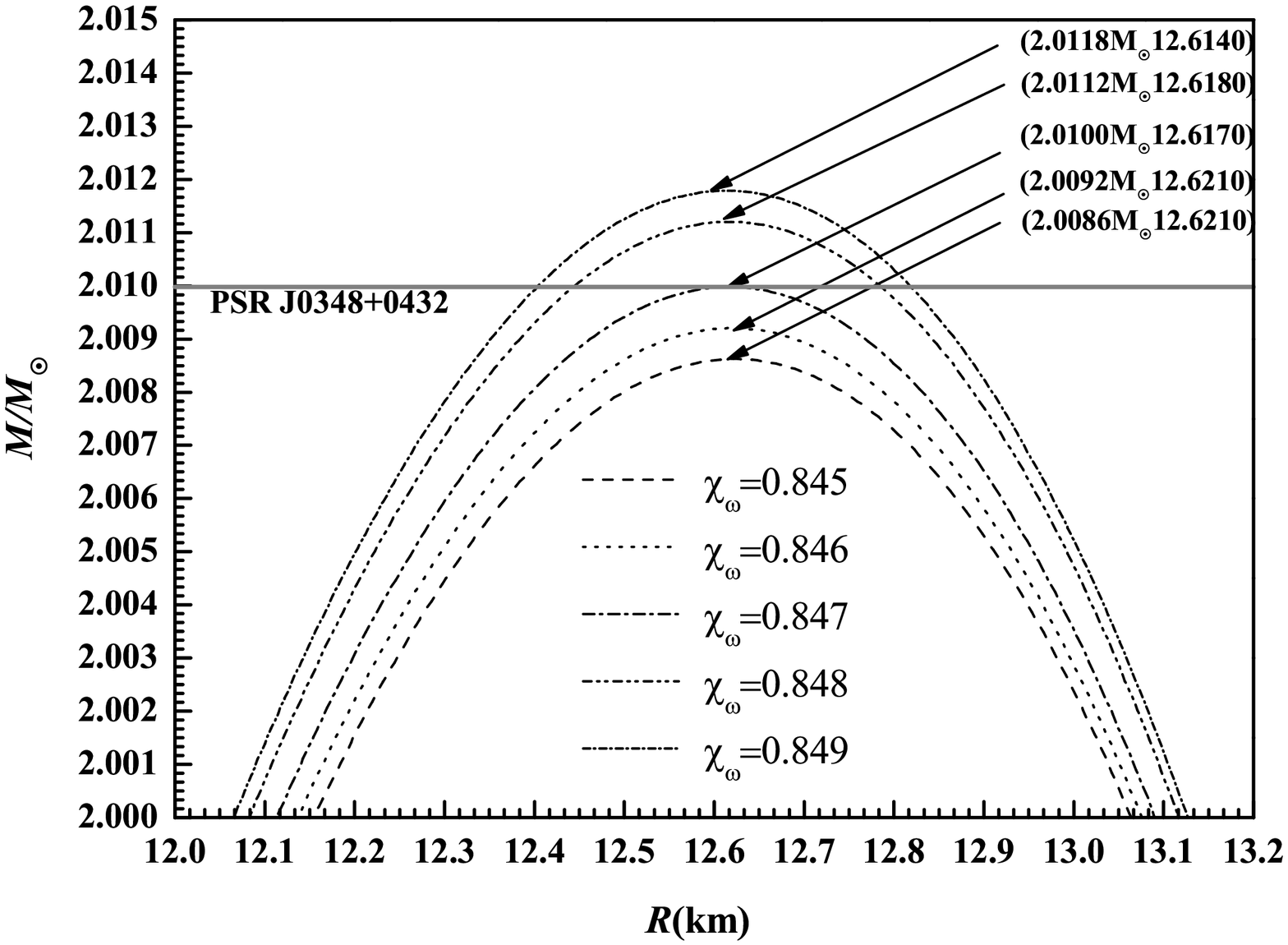}
\caption{Maximum neutron star mass of PNS as a function of radius in different $x_{\omega}$ without considering $\sigma^{*}$ and $\phi$.}
\label{figure3}
\end{center}
\end{figure}

Now, we calculate the mass of a protoneutron star without considering $\sigma^{*}$ and $\phi$. For these nascent neutron stars, the thermal effect should be considered in an approximately uniform entropy per baryon from 0 to 10 \cite{38}. The neutrino effect may allow to specify the star characteristics in the interior and we will discuss in our future work. In this work, we don't consider the neutrino concentrations and select the entropy per baryon to 1. The Fig.~\ref{figure1} gives the $x_{\omega}$ is 0.77 to 1.0, so we select the extreme value of the $x_{\omega}$=1.0, the maximum mass calculated is 1.9624$M_{\odot}$, which can not describe the PSR J0348+0348 whose mass is 2.0100$M_{\odot}$. So the GL97 may not give a perfect describe in maximum protoneutron stars.
Similarly, the Fig.~\ref{figure2} gives the $x_{\omega}$ is 0.76 to 1, we use the same method to select the value of the $x_{\omega}$ by GL85. $x_{\omega}$=1.0, 0.9, 0.8, 0.76, the maximum mass are 2.1076$M_{\odot}$, 2.0507$M_{\odot}$, 1.9691$M_{\odot}$, 1.9294$M_{\odot}$ respectively. Clearly, the $x_{\omega}$ among 0.8 and 0.9 maybe give the mass of 2.0100$M_{\odot}$. At first, we select $x_{\omega}$=0.85 and get the maximum mass is 2.0126$M_{\odot}$, which is bigger than 2.01$M_{\odot}$. We select $x_{\omega}$=0.84 and get the maximum mass is 2.0041$M_{\odot}$, which is smaller than 2.01$M_{\odot}$. Then, we select $x_{\omega}$=0.845 and get the maximum mass is 2.0086$M_{\odot}$. So we can constrict the value of $x_{\omega}$ between 0.845 and 0.85, $x_{\omega}$=0.846, 0.847, 0.848, 0.849, the maximum masses are 2.0092$M_{\odot}$, 2.0100$M_{\odot}$, 2.0112$M_{\odot}$, 2.0118$M_{\odot}$ respectively. The calculations are shown in Fig.~\ref{figure3} and Table~\ref{tab:table3}. The discussion finally give the hyperon coupling constants $x_{\omega\Xi}$=$x_{\omega\Sigma}$=$x_{\omega\Lambda}$=0.847, corresponding to $x_{\sigma\Xi}$=0.670,$x_{\sigma\Sigma}$=0.462,$x_{\sigma\Lambda}$=0.740, and $x_{\rho\Lambda}$=0,$x_{\rho\Sigma}$=2,$x_{\sigma\Xi}$=1, which we get the maximum mass of a PNS is 2.0100$M_{\odot}$ without considering $\sigma^{*}$ and $\phi$.

\begin{table}[htb]
\begin{center}
\caption{\label{tab:table3}
 Without $\sigma^{*}$ and $\phi$
}
\begin{ruledtabular}
\begin{tabular}{lcccccccc}
 $x_{\omega}$&1.0 &0.9 &0.8 &0.76 \\
M$_{\mathrm{max}}$   &2.1076M$_{\odot}$      &2.0507M$_{\odot}$   &1.9691M$_{\odot}$ &1.9294M$_{\odot}$\\
\hline
 $x_{\omega}$& &0.85 &0.84 &\\
M$_{\mathrm{max}}$   &      &2.0126M$_{\odot}$   &2.0041M$_{\odot}$  &\\
\hline
 $x_{\omega}$&0.845 &0.846 &$\mathbf{0.847}$ & 0.848 &0.849\\
M$_{\mathrm{max}}$   &2.0086M$_{\odot}$      &2.0092M$_{\odot}$  & $\mathbf{2.0100M_{\odot}}$ &2.0112M$_{\odot}$ &2.0018M$_{\odot}$\\
\end{tabular}
\end{ruledtabular}
\end{center}
\end{table}

\begin{figure}[!htb]
\begin{center}
\includegraphics[width=3.5in]{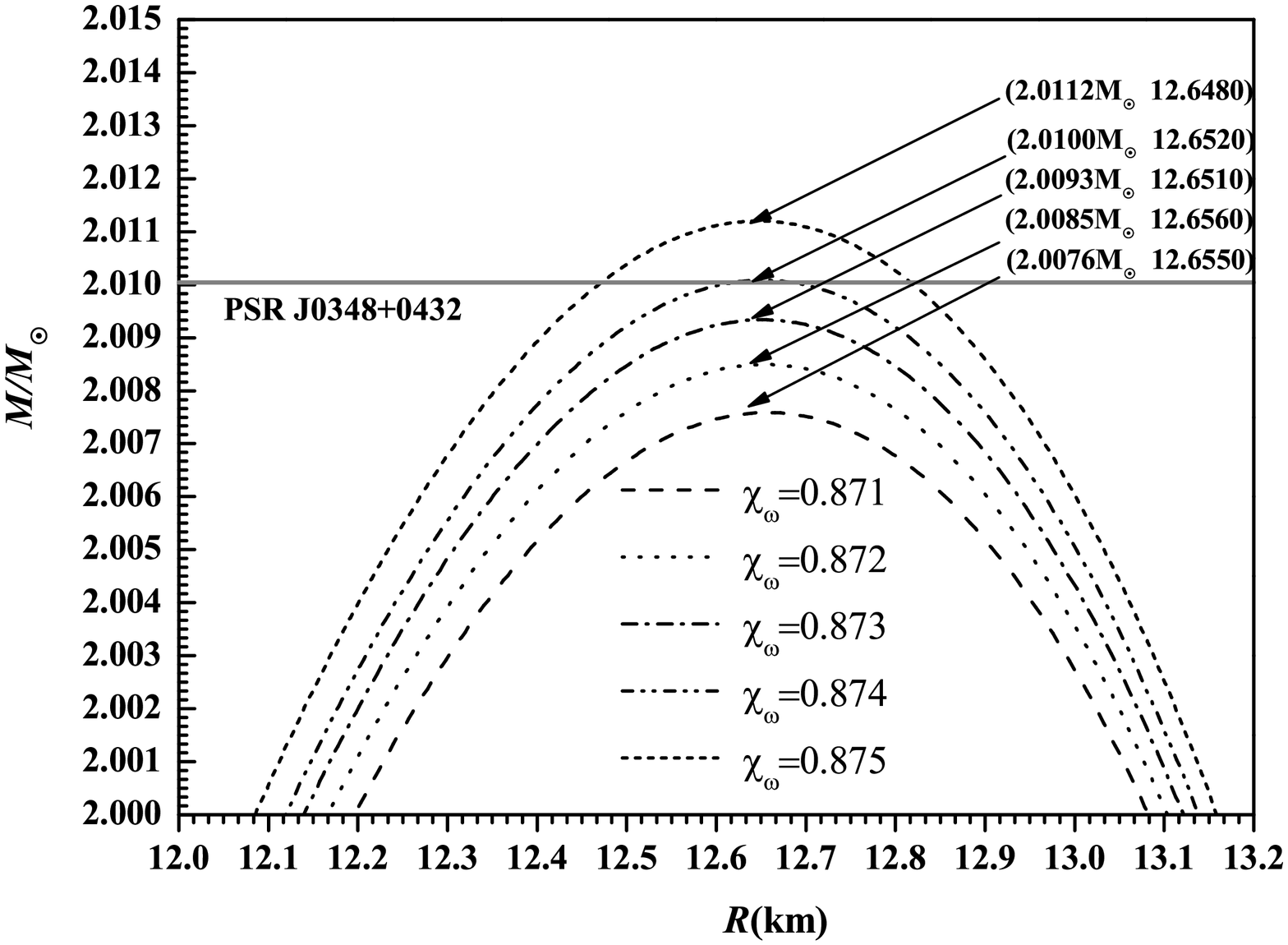}
\caption{Maximum neutron star mass of PNS as a function of radius in different $x_{\omega}$ with considering $\sigma^{*}$ and $\phi$.}
\label{figure4}
\end{center}
\end{figure}

When the $\sigma^{*}$ and $\phi$ mesons are took into account, our aim is also to get the maximum mass of a PNS corresponding to PSR J0348+0432. We select the value of $x_{\omega}$=1.0, 0.9, 0.8, 0.76 according to the Fig.~\ref{figure2}, the maximum mass are 2.0998$M_{\odot}$, 2.0449$M_{\odot}$, 1.9376$M_{\odot}$, 1.8919$M_{\odot}$ respectively and the $x_{\omega}$ among 0.8 and 0.9 maybe give the mass of 2.01$M_{\odot}$ by the same method above. At first, we select $x_{\omega}$=0.85 and get the maximum mass 1.9881$M_{\odot}$, which is smaller than 2.0100$M_{\odot}$. Then, we select $x_{\omega}$=0.86, 0.87, 0.88, 0.89 and get the maximum mass are 1.9975$M_{\odot}$, 2.0067$M_{\odot}$, 2.0156$M_{\odot}$, 2.0242$M_{\odot}$ respectively. On the basis of these results above, we select the value of $x_{\omega}$=0.875, which gives the maximum mass is 2.0112$M_{\odot}$ and is bigger than 2.0100$M_{\odot}$. So it is clear that we can constrict the value of $x_{\omega}$ between 0.87 and 0.875, the value of $x_{\omega}$=0.874, 0.873, 0.872, 0.871 give the maximum mass are 2.0100$M_{\odot}$, 2.0093$M_{\odot}$, 2.0085$M_{\odot}$, 2.0076$M_{\odot}$ respectively. The calculations are shown in Fig.~\ref{figure4} and Table~\ref{tab:table4}. The discussion finally give the hyperon coupling constants $x_{\omega\Xi}$=$x_{\omega\Sigma}$=$x_{\omega\Lambda}$=0.874, corresponding to $x_{\sigma\Xi}$=0.689,$x_{\sigma\Sigma}$=0.481,$x_{\sigma\Lambda}$=0.759, and $x_{\rho\Lambda}$=0,$x_{\rho\Sigma}$=2,$x_{\sigma\Xi}$=1, which we get the maximum mass of a PNS is 2.0100$M_{\odot}$ with considering $\sigma^{*}$ and $\phi$.

\begin{table}[htb]
\begin{center}
\caption{\label{tab:table4}
 With $\sigma^{*}$ and $\phi$
}
\begin{ruledtabular}
\begin{tabular}{lcccccccc}
 $x_{\omega}$&1.0 &0.9 &0.8 &0.76 \\
M$_{\mathrm{max}}$   &2.0098M$_{\odot}$      &2.0449M$_{\odot}$   &1.9376M$_{\odot}$ &1.8919M$_{\odot}$\\
\hline
 $x_{\omega}$&0.85 &0.86 &0.87&0.88&0.89\\
M$_{\mathrm{max}}$   &1.9881M$_{\odot}$      &1.9975M$_{\odot}$   &2.0067M$_{\odot}$  &2.0156M$_{\odot}$ &2.0242M$_{\odot}$  \\
\hline
 $x_{\omega}$&0.871 &0.872 &0.873 & $\mathbf{0.874}$ &0.875\\
M$_{\mathrm{max}}$   &2.0076M$_{\odot}$      &2.0085M$_{\odot}$  &2.0093M$_{\odot}$  &$\mathbf{2.0100M_{\odot}}$ &2.0112M$_{\odot}$\\
\end{tabular}
\end{ruledtabular}
\end{center}
\end{table}

 As a result, we give two sets of hyperon coupling constants to describe the PNS PSR J0348+0432 by GL85 with considering $\sigma^{*}$ and $\phi$ or not.  They are shown in Table~\ref{tab:table5}. The Fig.~\ref{figure3} shows that the PNS PSR J0348+0432 have the radius at 12.6170km without the $\sigma^{*}$ and $\phi$. When the $\sigma^{*}$ and $\phi$ are took into account, the result gives the radius is 12.6520km shown by the Fig.~\ref{figure4}, which is bigger than the radius of PNS PSR J0348+0432 without considering $\sigma^{*}$ and $\phi$.

 These results show that the $\sigma^{*}$ and $\phi$ are in favor of increasing the radius, and it means that the radius of the massive PNS PSR J0348+0432 with considering $\sigma^{*}$ and $\phi$ is bigger than that without considering $\sigma^{*}$ and $\phi$. But this transformation is so faint.

\begin{table}[htb]
\begin{center}
\caption{\label{tab:table5}
 Two sets of hyperon coupling constants describing the PNS PSR J0348+0432. CASE1 does not consider $\sigma^{*}$ and $\phi$ and CASE2 considers.
}
\begin{ruledtabular}
\begin{tabular}{lcccccccccccc}
 &$x_{\omega\Xi}$ &$x_{\omega\Sigma}$ & $x_{\omega\Lambda}$ & $x_{\sigma\Xi}$& $x_{\sigma\Sigma}$ & $x_{\sigma\Lambda}$ & $x_{\rho\Xi}$ & $x_{\rho\Sigma}$ &$x_{\rho\Lambda}$  \\
\hline
CASE1   &0.847    &0.847     &0.847   &0.670   &0.462  &0.740 & 1 &2 & 0\\
CASE2  &0.874 &0.874 &0.874 & 0.689 &0.481 & 0.759 &1 & 2 & 0 \\
\end{tabular}
\end{ruledtabular}
\end{center}
\end{table}

\subsection{COMPOSITION}

\begin{figure}[!htb]
\begin{center}
\includegraphics[width=3.5in]{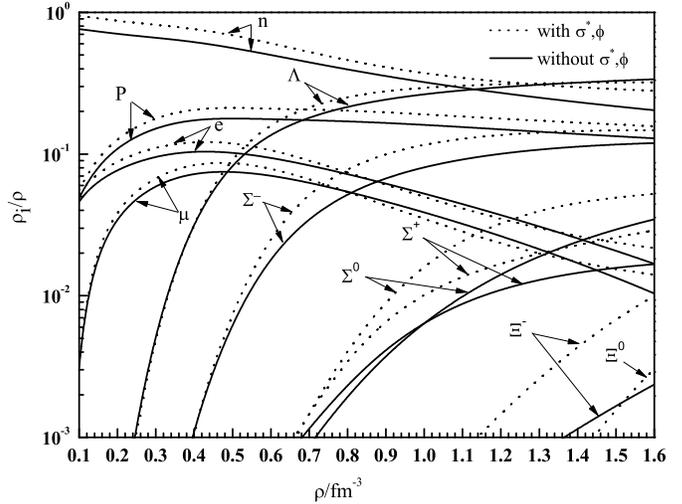}
\caption{Population of various particles in PNS PSR J0348+0432 with considering $\sigma^{*}$ and $\phi$ or not.}
\label{figure5}
\end{center}
\end{figure}

In a neutron star interior, some of the nucleons can be converted to hyperons which carry strangeness. The octet baryons comprise some of the least massive baryons which include the $\Lambda,\Sigma,\Xi$. These hyperons form a significant population of massive PNSs and indeed are dominant in the high density. The relative populations of various particles in PNS PSR J0348+0432 described in Fig.~\ref{figure5}. In this work, we select the GL85 parameter sets which give the unclear saturation density at 0.145$fm^{-3}$($\rho_{0}$), in neutron star interior the nucleons will convert to hyperon when the density exceed the unclear saturation density through the strong interaction. We can see the first hyperon to appear in the hadronic matter is $\Lambda$ at 0.248$fm^{-3}$(1.71$\rho_{0}$) no matter whether considering the $\sigma^{*}$ and $\phi$. The next hyperon is $\Sigma^{-}$ and $\Sigma^{-}$ appears almost simultaneously around 0.399 fm$^{-3}$ in both cases. But the density at where $\Sigma^{0}$ appears is about 0.661 fm$^{-3}$ with including $\sigma^{*}$ and $\phi$, 0.721 fm$^{-3}$ without including $\sigma^{*}$ and $\phi$, while the $\Sigma^{+}$ appears is about 0.670 fm$^{-3}$ with including $\sigma^{*}$ and $\phi$, 0.681 fm$^{-3}$ without including $\sigma^{*}$ and $\phi$. For hyperon $\Xi$, the density at where $\Xi^{-}$ appears is about 1.140 fm$^{-3}$ with including $\sigma^{*}$ and $\phi$, and 1.360 fm$^{-3}$ without including $\sigma^{*}$ and $\phi$.  With including $\sigma^{*}$ and $\phi$, the $\Xi^{0}$ appears is about 1.440 fm$^{-3}$. But without including $\sigma^{*}$ and $\phi$, the $\Xi^{0}$ appears at higher density which is not depicted in the figure.

These results show that in the context of considering $\sigma^{*}$ and $\phi$ in PNS PSR J0348+0432, the threshold for the appearance of the hyperons will be pushed forward comparing to without considering $\sigma^{*}$ and $\phi$.

\begin{figure}[!htb]
\begin{center}
\includegraphics[width=3.5in]{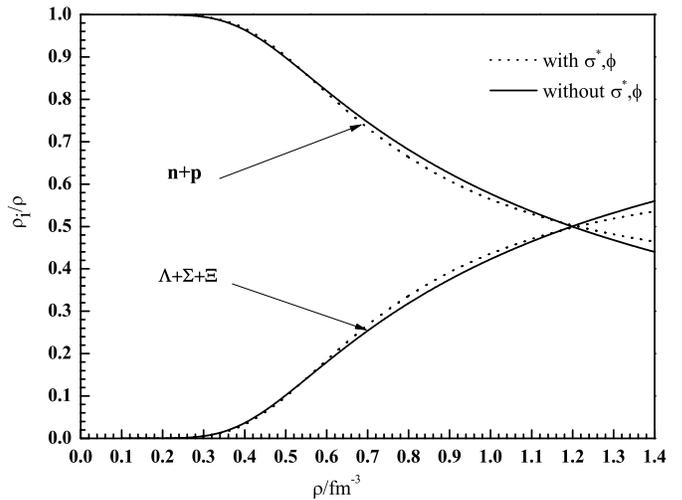}
\caption{Population of nucleon and hyperon as function of baryon density in PNS PSR J0348+0432 with considering $\sigma^{*}$ and $\phi$ or not.}
\label{figure6}
\end{center}
\end{figure}

 \begin{figure}[!htb]
\begin{center}
\includegraphics[width=3.5in]{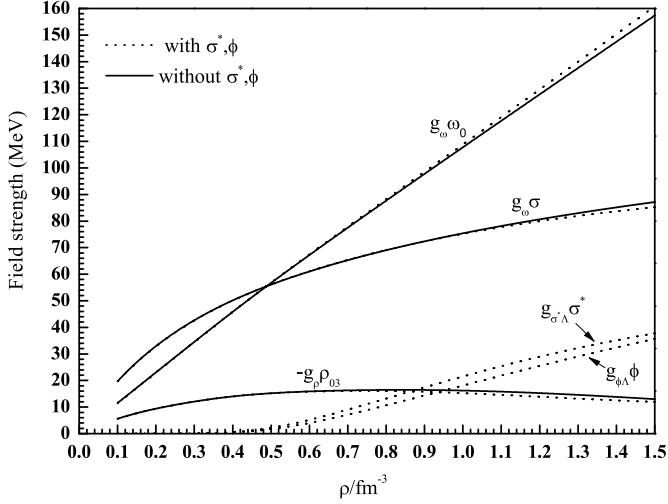}
\caption{Field strength of various mesons in PNS PSR J0348+0432 with considering $\sigma^{*}$ and $\phi$ or not.}
\label{figure7}
\end{center}
\end{figure}

The population of nucleons(n,p) and hyperons($\Lambda,\Sigma,\Xi$) as function of baryon density are shown in Fig.~\ref{figure6}. It clearly shows that when the hyperons appear, the numbers of nucleons will decrease. The numbers of hyperons will exceed the numbers of nucleons at 1.199$fm^{-3}$(8.27$\rho_{0}$) without $\sigma^{*}$ and $\phi$ included. We call this density as the transition density point which the hyperons start to play an important role in neutron star interior, therefore the canonical neutron star converts to the hyperon star. When we consider $\sigma^{*}$ and $\phi$ in the PNS PSR J0348+0432, the transition density point appears at 1.211$fm^{-3}$(8.35$\rho_{0}$).

The field strength of various mesons are shown in Fig.~\ref{figure7}. Here, when considering $\sigma^{*}$ and $\phi$ in PNS PSR J0348+0432, the $\omega$ gives the stronger field strength but $\sigma$ and $\rho$ give the weaker field strength. We also distinctly see from Fig.~\ref{figure7}, the field strength of $\sigma^{*}$ is larger than the field strength of $\phi$ and both increase with the baryon density. In relativistic mean field theory, the scalar meson $\sigma$ and $\sigma^{*}$ provides attraction, the vector meson $\omega$ and $\phi$ provides repulsion. When considering $\sigma^{*}$ and $\phi$, due to the attraction provided by $\sigma^{*}$ is larger than the repulsion provided by $\phi$, therefore, it will give the stiffer equation of state.

\subsection{MOMENT OF INERTIA  AND SURFACE GRAVITATIONAL REDSHIFT}
With the equations of state by resolving the TOV equation, we have got the mass and radius shown in Fig.~\ref{figure3} - Fig.~\ref{figure4}. With equations (14-19), the moment of inertia and surface gravitational redshift will be given.

\begin{figure}[!htb]
\begin{center}
\includegraphics[width=3.5in]{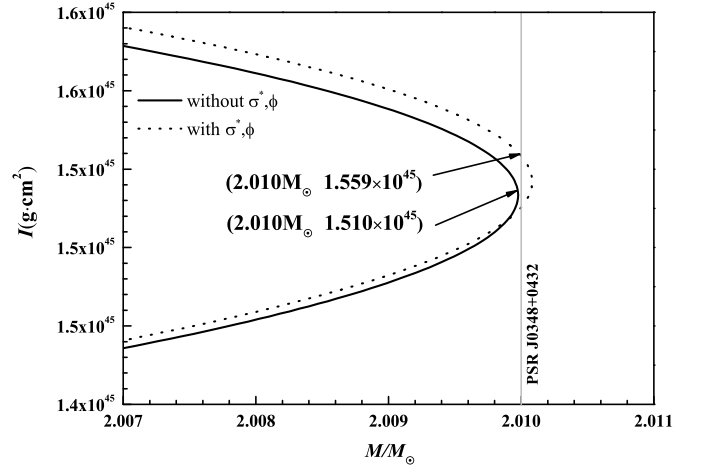}
\caption{The moment of inertia as a function of mass in different cases with considering $\sigma^{*}$ and $\phi$ or not. The shaded line corresponds to PNS PSR J0348+0432.}
\label{figure8}
\end{center}
\end{figure}

\begin{figure}[!htb]
\begin{center}
\includegraphics[width=3.5in]{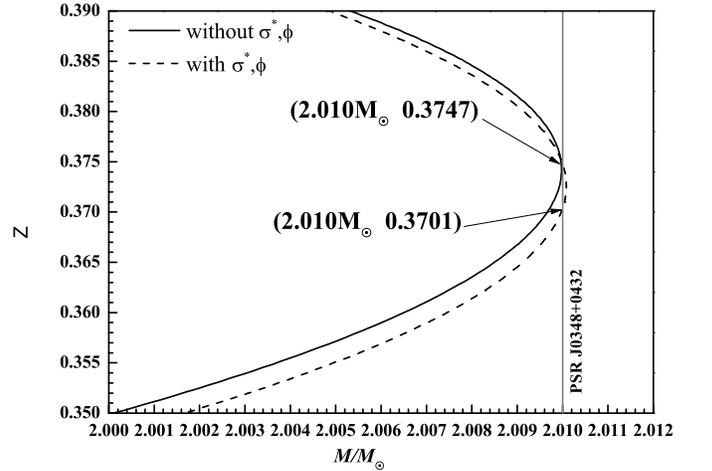}
\caption{The surface gravitational redshift as a function of mass in different cases with considering $\sigma^{*}$ and $\phi$ or not. The shaded line corresponds to PNS PSR J0348+0432.}
\label{figure9}
\end{center}
\end{figure}

\begin{table}[htb]
\begin{center}
\caption{\label{tab:table6}
 The properties of PNS PSR J0348+0432 with considering $\sigma^{*}$ and $\phi$ or not. CASE1 considers $\sigma^{*}$ and $\phi$ and CASE2 does not consider. $R$ is radius, $I$ is moment of inertia and $Z$ indicates surface gravitational redshift.
}
\begin{ruledtabular}
\begin{tabular}{lcccccccc}
 &M$_{\mathrm{max}}$ (M$_{\odot}$) &R (km) &I(g$\cdot$ cm$^{3}$) &Z \\
\hline
CASE1   &2.010     &12.6520     &1.559$\times10^{45}$ &0.3701\\
CASE2  &2.010      &12.6170   &1.510$\times10^{45}$ &0.3747\\
\end{tabular}
\end{ruledtabular}
\end{center}
\end{table}

The profile of moment of inertia in massive PNS with two cases is shown in Fig.~\ref{figure8}. We give the moment of inertia of the PNS corresponding to PSR J0348+0432 is $1.559\times10^{45}g.cm^{2}$with considering $\sigma^{*}$ and $\phi$ and is $1.510\times10^{45}g.cm^{2}$ without considering $\sigma^{*}$ and $\phi$. This points that the PNS PSR J0348+0432 with considering $\sigma^{*}$ and $\phi$ will increase the moment of inertia explained that the bigger radius by calculation above.

Likewise, the profile of surface gravitational redshift in massive PNS with two cases is shown in Fig.~\ref{figure9}. We give the gravitational redshift of the PNS corresponding to PSR J0348+0432 is $0.3701$ with considering $\sigma^{*}$ and $\phi$ and is $0.3747$ without considering $\sigma^{*}$ and $\phi$. This result tells us the PNS PSR J0348+0432 with considering $\sigma^{*}$ and $\phi$ will decrease the gravitational redshift, it is explained that the bigger radius will give the smaller gravitational redshift by the formula above.

All of the discussion calculated by us are listed in the Table~\ref{tab:table6}.

\section{Summary}
This paper discuss the influence of $\sigma^{*}$ and $\phi$ on the PNS PSR J0348+0432 in the framework of relativistic mean field theory comparing to the case without $\sigma^{*}$ and $\phi$. We restrict the value of $x_{\omega}$ at a narrow area in the nucleon coupling sets GL85 and GL97 and we also weed out the possibility that GL97 could describe the PNS PSR J0348+0432. In the context of GL85 sets, we give two sets of hyperon coupling constants to describe the PNS PSR J0348+0432 with considering $\sigma^{*}$ and $\phi$ or not. We use an entropy per baryon S=1 to emphasize thermal effect on PSR J0348+0432 in this work and study the different effects between considering $\sigma^{*}$,$\phi$ and without considering them in the PNS PSR J0348+0432. These effects include the particles number distribution, the mass-radius relation, the moment of inertia and surface gravitational redshift. We find that the PNS PSR J0348+0432 with $\sigma^{*}$ and $\phi$ will pushed forward the threshold for the appearance of the hyperons. We give the mass-radius relations are ($2.010M_{\odot}$,12.6520km) and ($2.010M_{\odot}$,12.6170km) corresponding to the PNS PSR J0348+0432 with and without considering $\sigma^{*}$ and $\phi$ respectively. It means that the $\sigma^{*}$ and $\phi$ are in favor of increasing radius. The moment of inertia corresponding to the PNS PSR J0348+0432 are ($2.010M_{\odot}$,1.510$\times 10^{45}$g.cm$^{2}$) and $(2.010M_{\odot}$,1.559$\times 10^{45}$g.cm$^{2})$ respectively, the surface gravitational redshift corresponding to the PNS PSR J0348+0432 are ($2.010M_{\odot}$,0.3747) and ($2.010M_{\odot}$,0.3701) respectively. These calculation tell us that considering the $\sigma^{*}$ and $\phi$ in the PNS PSR J0348+0432, the redshift will decrease while the moment of inertia will increase, but there is no discernible difference between the changes. These conclusions point that the $\sigma^{*}$ and $\phi$ give very little influence on PNS PSR J0348+0432, it means that the restriction of $\sigma^{*}$ and $\phi$ on the interactions between baryons play a minor role in PNS core.

Maybe our work still remains some shortcomings and gives some rudimentary discussions, we will make up for these in the future works.
\begin{acknowledgments}
This work was supported by National Natural Science Foundation of China (Grant No.11535004 and No.11761161001) and by the National Major State Basic Research and Development Program of China (Grant No.2016YFE0129300)
\end{acknowledgments}

\bibliography{apssamp}

\end{document}